\documentclass[11pt]{article}

\usepackage{graphicx}
\usepackage{floatflt}
\newcommand {\re}{{\rm e}}\newcommand {\ri}{{\rm i}}
\newcommand {\rK}{{\rm K}}
\newcommand {\rB}{{\rm B}}\newcommand {\rD}{{\rm D}}
\newcommand {\rS}{{\rm S}}\newcommand {\rL}{{\rm L}}
\newcommand {\rKq}{{\rm\overline K}}
\newcommand {\rBq}{{\rm\overline B}}\newcommand {\rDq}{{\rm\overline D}}

\newcommand {\rBz}{{\rm B^0}} \newcommand {\rBqz}{{\overline{\rm B}}{}^0}
\newcommand {\rBs}{{\rm B_s}} \newcommand {\rBqs}{\rm{\overline B}{}_s}
\newcommand {\rDz}{{\rm D^0}} \newcommand {\rDqz}{{\overline{\rm D}}{}^0}
\newcommand {\rKz}{{\rm K^0}} \newcommand {\rKqz}{{\overline{\rm K}}{}^0}
\newcommand {\rMz}{{\rm M^0}} \newcommand {\rMqz}{{\overline{\rm M}}{}^0}

\newcommand {\Begeq}{\begin{equation}}
\newcommand {\Endeq}{\end{equation}}
\newcommand {\bEa}{\begin{eqnarray}}
\newcommand {\eEa}{\end{eqnarray}}

\newcommand {\rb}[1]{\raisebox{-0.3ex}[-0.1ex]{#1}}

\begin{document}

\begin{center}
{\bf{\large{
Demonstration of $\rKz\rKqz$, $\rBz\rBqz$, and 
$\rDz\rDqz$ Transitions\\ with a Pair of Coupled Pendula
}}}\\[8mm]
Klaus R.\ Schubert\\ Institut f\"ur Kern- und Teilchenphysik, Technische
Universit\"at Dresden\\[3mm]
and J\"urgen Stiewe\\ Kirchhoff-Institut f\"ur Physik,
Universit\"at Heidelberg
\\[8mm]\end{center}

\noindent
Abstract: A setup of two coupled and damped pendula is used to demonstrate 
the main features of transitions between the neutral mesons $\rKz$, $\rDz$,
$\rBz$ and their antiparticles, including CP violation in the $\rKz$ system.
The transitions are described by two-state 
Schr\"odinger equations. Since the real parts of their solutions obey the same
differential equations as the pendula coordinates, the pendulum motions 
can be used
to represent the meson transitions. Video clips of the motions are
attached as supplementary material.\\

\begin{figure}[h]\begin{center}
\includegraphics[width=5.5cm]{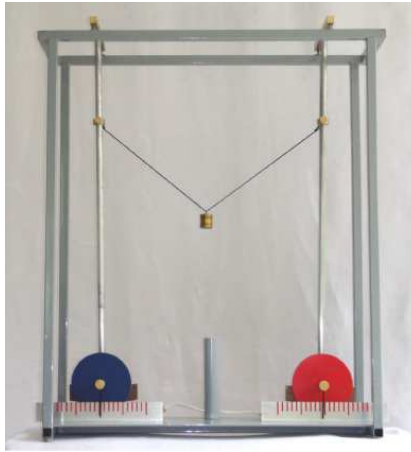}\end{center} 
\caption[ ]{Photograph of the demonstration pendula. \label{Fig:1.1}}
\end{figure}
\vspace{1mm}

\noindent
1. Introduction\\

\noindent
We describe two coupled pendula which
we built, inspired by a presentation of B.\ Winstein \cite{1988-Winstein}, 
in 1987 when transitions between $\rBz$ and $\rBqz$ mesons
had been discovered \cite{1987-ARGUS-Bmixing}. Since then, the pendula
have been presented many times for visualizing the evolution of
states $\psi_1 \rMz + \psi_2 \rMqz$ where $\rMz$ is a $\rKz$, $\rDz$, or
$\rm \rBz$ meson, and
$\rMqz$ is its respective antiparticle. A meson is an elementary particle 
consisting of one quark and one antiquark. It, therefore, belongs neither to
matter nor to antimatter. Some mesons are their own
antiparticle like the photon, an example is the neutral $\pi$-meson $\pi^0$. 
The $\rKz$ meson
and its antiparticle $\rKqz$ are different particles, they differ in their
properties and their quark content. The phenomenon that a $\rKz$ may 
transform itself spontanuously
into a $\rKqz$, called $\rKz\rKqz$ transition, is a non-trivial but now
well-understood consequence of the weak interaction. 
In a world with six quark flavors,
where one quark (top) is so heavy that it decays 
before it can form a meson, the above-mentioned $\rMz\rMqz$ pairs 
and $\rm B_s {\overline B}{}_s$
are the only four which show transitions $\rMz\leftrightarrow\rMqz$.
All four transition types have been observed in experiments
at particle accelerators.\\

\noindent 
In these experiments studies of the discrete symmeties under the
operations C, P, CP, T, and CPT have played an important role, where C 
denotes the exchange of particles and antiparticles like $\rKz\to\rKqz$, 
P the parity operation $\vec r\to - \vec r$, and T the time-reversal
operation ($t\to -t$ and exchanging ingoing and outgoing states).\\ 

\begin{table}[h]
Table 1: Transition parameters and properties of the four two-state 
systems $\rMz\rMqz$ \cite{2010-PDG,2011-HFAG}.
The parameters in the first five lines are defined 
in the text. The parameter
$|p/q|$  decribes CP violation in the transitions. If
no error is given, the precision of the parameter is higher than the
number of digits given. 
The last line gives the property combinations 
of the two fundamental states for each system. These states are Long 
($L$)- or Short ($S$)-living, heavy ($h$) or light ($\ell$), CP-even (+)
or CP-odd (-); there are no first-principle relations between the three
properties. 
\begin{center}
\vspace{3mm}\begin{tabular}{ccccc}
\hline\hline\vspace*{1mm}
& \hspace{6mm}\rb{$\rKz\rKqz$}\hspace{6mm} & \rb{$\rDz\rDqz$} & 
\rb{$\rBz\rBqz$} & \rb{$\rBs\rBqs$}\\ 
\hline
$\mu$ [MeV] & 498 & 1865 & 5279 & 5366\\
$\Gamma~[\rm s^{-1}]$ & $5.59\cdot 10^9$ & $2.44\cdot 10^{12}$ & 
   $6.6\cdot 10^{11}$ & $6.8\cdot 10^{11}$ \\
$|\Delta m| [\rm s^{-1}]$ & $5.3\cdot 10^9$ & $(1.5\pm 0.5)\cdot 10^{10}$ &
   $5.1\cdot 10^{11}$ & $1.8\cdot 10^{13}$ \\
$\Delta\Gamma~[\rm s^{-1}]$ & $1.115\cdot 10^{10}$ & $(3.9\pm 0.6)\cdot 10^{10}$ & 
   $<~3\cdot 10^{10}$ & $(6\pm 2)\cdot 10^{10}$ \\ 
$\chi$ & 0.4986 & $(5.1\pm 1.6)\cdot 10^{-5}$ & 0.19 & 0.4993 \\ 
$|p/q|-1$ & $3.3\cdot 10^{-3}$ & $0.1 \pm 0.2$ & $(-2.4\pm 2.3)\cdot 10^{-3}$ 
    &  $(-4\pm 3)\cdot 10^{-3}$ \\ \hline\vspace*{1mm}
& \rb{$\ell = + = S$} & \rb{$h = + = S$} & \rb{$h = ? = ?$} & \rb{$? = + = S$}\\   
\hline\hline\end{tabular}\end{center} 
\end{table} 

\noindent
 Transitions between $\rKz$ and
$\rKqz$ mesons were predicted by M.\ Gell-Mann and A.\ Pais
\cite{1955-GellMannPais} in 1955. First observations 1958
in nuclear emulsion \cite{1958-BaldoCeolin} and in a hydrogen bubble
chamber \cite{1959-Crawford} were based on the appearance of $\rKqz$
mesons in a $\rKz$ beam. These and all later $\rKz\rKqz$
observations are successfully described by a two-state Schr\"odinger equation. 
Many superpositions 
$\psi_1\rKz + \psi_2\rKqz$ are solutions of the equation, but only two
of them have an exponential decay law. They are called $\rKz_\rS$ with
a decay rate $\Gamma_\rS$ and $\rKz_\rL$ with $\Gamma_\rL\ll\Gamma_\rS$.
These states have also different masses, $m_\rS$ and $m_\rL$ respectively.
Because of (weakly broken) CP symmetry,
the states are also (approximate) CP eigenstates; CP($\rKz_\rS)=+1$ and 
CP($\rKz_\rL)=-1$. The latest experimental results for
$\mu=(m_\rS+m_\rL)/2$,  $\Delta m = m_\rS - m_\rL$, 
$\Gamma=(\Gamma_\rS+\Gamma_\rL)/2$,
and $\Delta\Gamma =\Gamma_\rS - \Gamma_\rL$ are given in Table 1.
 The transition strength is given by the parameter 
$\chi(\rMz)$,
\Begeq
\rm\chi(\rMz)=fraction~of~all~\rMz~decaying~as~\rMqz~=
\frac{4(\Delta m)^2+(\Delta\Gamma)^2}{8\Gamma^2+8(\Delta m)^2}~,~~\chi(\rKz)
= 0.4986.
\Endeq

\noindent
$\rBz\rBqz$ transitions were first observed 1987 at DESY
\cite{1987-ARGUS-Bmixing}. Present values for $\mu$,
$\Delta m$, and $\Gamma$ are given in Table 1;
for $\Delta\Gamma$ we have only an upper limit, and
\Begeq
\rm\chi(\rBz)=0.19.
\Endeq
First evidence for $\rBs\rBqs$ transitions, more precisely for the sum of
$\rBs\leftrightarrow\rBqs$ and $\rBz\leftrightarrow\rBqz$, was reported
by the UA1 group \cite{1987-UA1} in 1987. A significant determination of
$\Delta m(\rBs)$, however, could be reached only in 2006 at Fermilab
\cite{2006-CDF-Delta(mBs), 2006-D0-Delta(mBs)}; we now know that
\Begeq
\rm\chi(\rBs)=0.4993.
\Endeq 
$\rDz\rDqz$ transitions were discovered in 2007 by the experiments
BABAR \cite{2007-BABAR-Dmixing} and BELLE \cite{2007-BELLE-Dmixing}. 
Additional data in the
last four years lead to the results in Table 1, and
\Begeq
\rm\chi(\rDz)=5\cdot 10^{-5}.
\Endeq

\noindent
The study of $\rMz\rMqz$
transitions has played an important role in the progress of particle physics, 
especially in understanding the weak interaction. A summary would be
beyond the scope of our article, we only mention
the 1964 discovery of CP-symmetry breaking in $\rKz\rKqz$ transitions
\cite{1964-FitchCronin} leading to the Physics Nobel Prize in 1980. 
The mere observation that $\Delta m$, $\Gamma$, and $\Delta\Gamma$
are very small compared to $\mu$ for each meson pair shows that the
transitions are produced by weak interactions. This ensures two relevant facts:
First, the evolution of the states $\Psi=(\psi_1,\psi_2)$ for each pair
is given by a linear differential equation, the
Schr\"odinger equation $\ri~\partial\Psi/\partial t ={\bf M}~\Psi$ 
with a $2\times 2$ complex matrix {\bf M}; each pair
has a different set of eight real constants in $\bf M$. 
Second, the real parts $\Re(\psi_1)$ and $\Re(\psi_2)$
contain the same information as the complex solutions $\Psi$.
The latter is the basis for the main concern of the article: The
evolutions of  $\rMz\rMqz$ transitions may be represented by motions
of coupled pedula.\\

\noindent
Owing to arbitrary phases of
the states $\rMz$ and $\rMqz$, only seven of the eight constants in $\bf M$ have 
a physical meaning. The most general solution $\Psi$ has seven parameters which
follow unambiguously from the seven constants in $\bf M$. Discrete
symmetries reduce the number of constants and parameters 
\cite{BrancoLavouraSilva}; CPT symmetry leaves five, T symmetry leaves
six, and the combination of CPT and T, which also means CP symmetry,
leaves four. In Section 4, we discuss the most general CP-symmetric 
equation and its solutions, in Appendix 2 
the most general CPT-symmetric case. 
The motions of our coupled pendula represent 
the real parts of $\Psi$, they obey the same differential equations
as the deflection angles $\phi_1$
and $\phi_2$ of the pendula. The pendula are described in Section 2.
In Sections 3 and 4,
we discuss the Schr\"odinger-equation solutions 
for a stable particle, a decaying particle, and a quasi-stable two-particle
system, corresponding to the motions of 
an undamped pendulum, a damped pendulum, and a pair of
coupled but undamped pendula, respectively.\\

\noindent
The motions are shown in a series of video clips
which are attached as supplementary material
\cite{links}. In Section 5,
we discuss the most general CP-symmetric two-state Schr\"odinger equation 
and the corresponding equations of motion. 
In Section 6, we present $\rKz\rKqz$, $\rBz\rBqz$, and $\rDz\rDqz$ 
transitions with the same pendulum setup by translating the three 
parameter sets in $\bf M$ into the corresponding
parameters for coupling and damping the pendula.
In Section 7 we discuss CP-symmetry breaking in transitions
$\rKz\leftrightarrow\rKqz$. We choose an asymmetric pendulum setup
for demonstrating the main feature of this CP violation,
$\rKz_\rL=p~\rKz+q~\rKqz$ with $|p/q|>1$. However, the presented
setup has an equation of motion which does not correspond to a 
CPT-symmetric and T-violating Schr\"odinger equation.\\[3mm]

\noindent 
2. Description of the Pendulum Pair\\

\noindent
Fig.~\ref{Fig:1.1} shows a photograph of our setup. The two mechanically 
identical pendula, differing only by colour, are suspended in a steel 
housing and can swing freely, nearly frictionless.
The coupling between the pendula is established through a thin thread 
which connects the pendulum rods and can carry weights of different 
magnitudes as to strengthen or loosen the coupling. Also the length of the
thread and its connection points to the rods are adjustable. Each pendulum 
carries a copper sheet which can slide in the opening of an electromagnet 
in order to introduce individual damping by
exciting eddy currents in the copper sheets.
Damping in the coupling is also possible, see below.
A technical drawing of steel housing and pendula, including the major 
measures, is shown in Fig.~\ref{Fig:3.2}. Detailed construction drawings
with all measures and a 3D drawing  are
available in the supplementary material \cite{links}.\\

\begin{figure}[h]\begin{center}
\includegraphics[width=9cm]{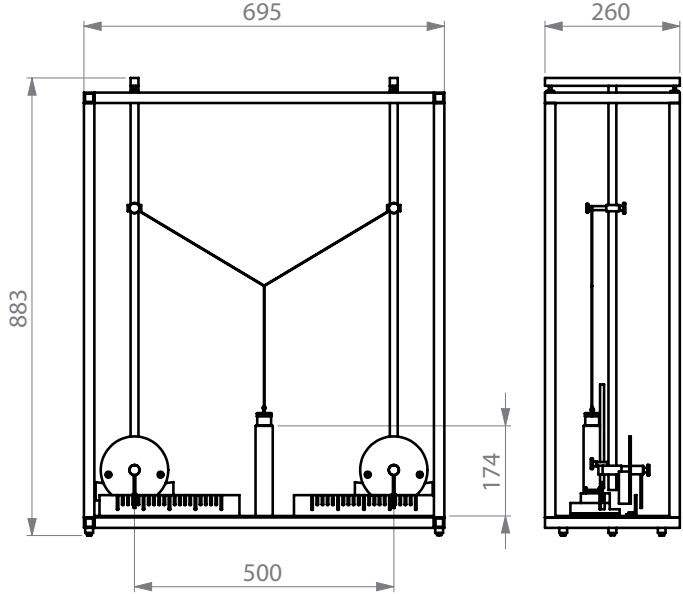}\end{center} 
\caption{Pendula and frame with major measures in mm.
\label{Fig:3.2}} 
\end{figure}

\noindent
The frame which houses the pendula is made of steel tubes with a 
square cross section, the ground plate consists of an aluminium sheet. 
The two long tubes on the upper side of the frame carry two half-globular 
pits with 15 mm diameter, made of case-hardened steel. 
The pits
serve as abutments for the steel pins on the pendula, see below. 
The distance 
between the two pendula is 500 mm. For measuring the amplitudes of
the pendula, two scales with spacings of 1 cm 
each are attached to the ground plate in variable positions.\\

\noindent
%
Each of the two pendula consists of a round aluminium rod on the 
upper side of which a brass traverse bar is fixed giving
the upper side of the pendulum a "T"-shaped appearence. Below the traverse 
bar, two pins of case-hardened steel are fixed
fitting
into the above-mentioned steel pits. Their tips represent 
the centres of the pendulum movements. The rotation axes of the pendula 
are defined by the points where the pins touch the pits.
The pendulum bobs are thin circular aluminium plates 
%
screwed to the pendulum rods 
whose distances $\ell$ from the pendulum
axes can be varied; for the demonstrations we use $\ell =$ 710 mm. Each plate
carries on its back a brass weight of 0.6 kg adapted to the plate shape.\\

\noindent
The two ends of the thread 
connecting the pendula
can be fixed at the pendulum rods at 
different distances from the rotation axes so that different lever arms 
can be chosen. In our standard demonstration, corresponding to the 
attached
video clips, the thread is 100 mm longer than the pendulum distance, the two
lever arms are 180 mm long, and the weights are 150, 250, and 12 grams for
$\rK\rKq$, $\rB\rBq$, and $\rD\rDq$ transitions, respectively.\\

\noindent   
Screwed to the lower end of each pendulum rod is a brass bar of circular 
cross section with a copper sheet soldered to its lower end. The copper 
sheet is 3 mm thick and has the shape of a circle segment following 
the pendulum bob\rq s oscillation trajectory. It fits in the opening of an 
electromagnet fixed to the aluminium ground plate of the pendulum frame.
The coil is driven by an AC voltage of $U_{\rm eff}= 18$ V and 
produces a field with an amplitude
of 0.10 Tesla exciting eddy currents in the copper 
sheet which damp the oscillation of each pendulum separately.\\

\noindent
Damping in the coupling - a salient point - is realized in the folowing
way:
The brass weight which is 
used to establish the coupling between the pendula in the $\rK\rKq$
demonstrations (Pos.\ 38 in the construction drawings \cite{links})
can be suspended on a separate thread such that it can dive 
into a brass cylinder which is
fixed vertically to the ground plate. The inner diameter of the cylinder 
corresponds to the weight\rq s diameter. A narrow bore runs through the brass 
weight in parallel to its vertical axis. The
vertical up- and down movements of the weight, following the pendulum 
swing, press air through the bore which flows in a laminar way, due to the 
slow movement. The air flow can be adjusted by a screw rectangular to 
the bore. \\[3mm] 

\noindent
3. Stable and Unstable Particles\\

\noindent
Using $\hbar =c=1$, the Schr\"odinger equation of a stable particle 
with mass $m$ in its rest frame is given by
\Begeq
\ri~ \partial\psi/\partial t = m\cdot\psi ~,\label{Eq:4.1} 
\Endeq
where $t$ is its eigentime. The complex solution, 
\Begeq
\psi = \Re+\ri\Im = \re^{-\ri m t}~, 
\Endeq
has two components which carry the same information as $\psi$,
the mass $m$ and
\Begeq
|\psi|^2= 2 \langle \cos^2 mt\rangle = 2 \langle \sin^2 mt\rangle =1,  
\Endeq
where $\langle ~ \rangle$ means averaging over times $T\gg 1/m$.
With in addition $T \ll 1/\Gamma$ and  $T\ll 1/\Delta m$, the property
$\langle \Re_i^2\rangle =|\psi_i|^2/2$, $i=1,2$ holds for all cases 
discussed in this article, as shown in Appendix 1.
The real part of $\psi$ obeys
\Begeq
\ddot\Re = -m^2~ \Re~.
\Endeq
In small-angle approximation, the equation of motion for a mathematical
pendulum with
length $\ell$ and acceleration $g$ is the same when $m^2=g/\ell$.
Video01 \cite{links} shows this motion with the oscillation period 
$2\pi/m = 1.69$ s. The stable particle is represented by the 
undamped pendulum.\\

\noindent
The effective Schr\"odinger equation of an unstable particle,
effective since it does not include the final states
into which the particle may decay, is given by
\Begeq
\ri~ \partial\psi/\partial t = (m-\ri\Gamma/2) \cdot\psi~,\label{Eq:5.1} 
\Endeq
where the total decay rate $\Gamma$ is the sum of the partial rates into
all final states. If the particle decays only weakly with $\Gamma\ll m$,
Eq.~\ref{Eq:5.1} is a very good
approximation \cite{1961-JacobSachs}. Its solution is
\Begeq
\psi = \Re+\ri\Im = \re^{-\ri m t}\cdot \re^{-\Gamma t/2},
\Endeq
where $\Re$ obeys 
\Begeq
\ddot\Re = -(m^2+\Gamma^2/4)\Re-\Gamma\dot\Re~,
\Endeq
the equation of motion for a pendulum with velocity-proportional damping.
Again $\psi$ and $\Re$ carry the same information, 
$m$ and 
\Begeq
|\psi|^2= 2 \langle \Re^2 \rangle= \re^{-\Gamma t}~.
\Endeq
The motion with $1/\Gamma = 7$ s is shown in Vido02 \cite{links}; its  
frequency is 
insignificantly smaller than that of the undamped pendulum. The time-averaged
amplitude square represents the probability that the particle has not yet 
decayed. \\[3mm]

\noindent
4. Transitions in a Symmetric Pair of Stable Particles\\

\noindent
The next introductory example, two coupled same-length pendula 
with negligeable damping, 
has no analogy in meson physics but corresponds
to the two states of an ammonia molecule as discussed e.\ g.\ by
R.\ Feynman \cite{FeynmanLectures}. The Schr\"odinger equation for a
two-state system of coupled particles with negligeable damping can be written as
\Begeq
\ri~\frac{\partial\Psi}{\partial t }={\bf M}~\Psi=\left(
\begin{array}{cc}m+k&-k\\-k&m+k\end{array}\right)\Psi~.\label{Eq:6.1}
\Endeq

\begin{figure}[h]\begin{center}
\includegraphics[width=6cm]{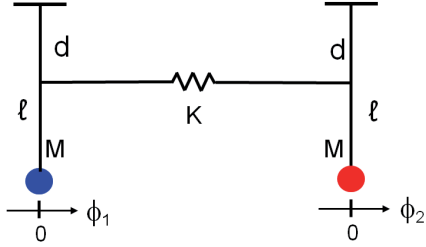}\end{center}
\caption[ ]{Coupled mathematical pendula without damping; 
coordinates are the angles $\phi_1$ and $\phi_2$, parameters are
mass $M$, pendulum length $\ell$, gravitational acceleration $g$,
and the two parameters for the coupling torque,
lever arm $d$ and proportionality constant $K$. \label{Fig:6.1}}
\end{figure}

\noindent
The linear combinations $\psi_1+\psi_2$ and $\psi_1-\psi_2$ obey uncoupled 
equations, leading to 
\Begeq
(\ddot\Re_1+\ddot\Re_2)=-m^2\cdot (\Re_1+\Re_2)~,
~~ (\ddot\Re_1-\ddot\Re_2)=-(m+2k)^2\cdot (\Re_1-\Re_2)\label{Eq:6.4}
\Endeq
and consequently to
\Begeq
\left(\begin{array}{c}\ddot\Re_1\\ \ddot\Re_2\end{array}\right)
=-\left(\begin{array}{cc}m^2+2mk+2k^2&-2mk-2k^2\\ 
-2mk-2k^2&m^2+2mk+2k^2\end{array}\right)
\left(\begin{array}{c}\Re_1\\ \Re_2\end{array}\right)\label{Eq:6.2}
\Endeq
for the real parts $\Re_1$ and $\Re_2$ of  $\psi_1$ and $\psi_2$.
Our mechanical analogon is the pendulum pair in Fig.~\ref{Fig:6.1}.
For the left-side pendulum, the coupling force is $F_{12}=
-K(d\cdot\phi_1 - d\cdot\phi_2)$, for
the other one $F_{21}=-F_{12}$, and the torques are $F_{\rm ij}\cdot d$.
The equation of rotary motion is
\Begeq
M\ell^2\left(\begin{array}{c}\ddot\phi_1\\ \ddot\phi_2\end{array}\right)
=-\left(\begin{array}{cc}Mg\ell+Kd^2&-Kd^2\\ 
-Kd^2&Mg\ell+Kd^2\end{array}\right)
\left(\begin{array}{c}\phi_1\\ \phi_2\end{array}\right)~.\label{Eq:6.3}
\Endeq
The differential equations in Eqs.~\ref{Eq:6.2} 
and \ref{Eq:6.3} are identical when $m^2=g/\ell$ and $2mk+2k^2 =Kd^2/(M\ell^2)$,
i.\ e.\ these pendula motions represent undamped transitions between two
quantum-mechanical states. 
With $\phi_i=\Re_i$ the solutions of Eqs.~\ref{Eq:6.4} are
\Begeq
\phi_1+\phi_2=\alpha\cdot\cos(mt-\beta)~,~~
\phi_1-\phi_2=\gamma\cdot\cos[(m+2k)t-\delta]~,
\Endeq
and their sum is the most general pendulum motion with four 
adjustable parameters to be
fixed by the conditions at $t=0$. We now show four
special solutions:\\

\noindent
i) The parallel fundamental mode in Video03 \cite{links} with 
$\gamma=\beta=0$, 
\Begeq
\phi_1=\phi_2=\phi_0\cdot\cos m t~.
\Endeq
ii) The antiparallel fundamental mode in Video04 \cite{links} with 
$\alpha=\delta=0$,
\Begeq
\phi_1=-\phi_2=\phi_0\cdot\cos (m+2k) t~.
\Endeq
Its frequency is larger than in case (i); the ``mass
difference" $2k$ is given by the matrix element
$M_{12}$ in Eq.~\ref{Eq:6.1}. 
The initial conditions for modes i) and ii) are the only ones which lead to
time-independent amplitudes for $\phi_1$ and $\phi_2$. All other conditions
lead to oscillation modes with beats, i.\ e.\ with varying amplitudes.\\

\noindent
iii) ``Blue starts and Red is at rest" in Video05 \cite{links} with
$\phi_1(0)=\phi_0$, 
$\phi_2(0)=\dot\phi_1(0)=\dot\phi_2(0)=0$ leading to
\bEa
\phi_1(t)=\phi_0\cdot[\cos(\mu-k)t+\cos(\mu+k)t]/2=\phi_0\cdot\cos kt
\cdot\cos\mu t~,\nonumber \\
\phi_2(t)=\phi_0\cdot[\cos(\mu-k)t-\cos(\mu+k)t]/2=\phi_0\cdot\sin kt
\cdot\sin\mu t~,
\eEa
where $\mu = m+k$. Our demonstration shows the oscillation 
period in between that of modes i) and ii) and the
beat period of $2\pi/k =57$ s.\\

\noindent
iv) ``Red starts and Blue is at rest" in Video06 \cite{links} with
$\phi_2(0)=\phi_0$, 
$\phi_1(0)=\dot\phi_1(0)=\dot\phi_2(0)=0$, leading to
$\phi_1(t)=\phi_2$ of motion (iii) and
$\phi_2(t)=\phi_1$ of motion (iii). The two motions are perfect mirror
images of each other; this symmetry
corresponds to CP symmetry in the meson transitions.\\[3mm]

\noindent
5. Transitions in a CP-Symmetric Meson Pair\\

\noindent
The most general Schr\"odinger equation 
for the evolution of the state 
$\psi_1~\rMz~+~\psi_2~\rMqz$ 
including transitions between $\rMz$ and $\rMqz$ is
\Begeq
\ri~\frac{\partial}{\partial t }\left(\begin{array}{c}\psi_1\\
\psi_2\end{array}\right)=
\left[\left(\begin{array}{cc}m_{11}&m_{12}\\ m_{12}^*&m_{22}\end{array}\right)
-\frac{\ri}{2}\left(\begin{array}{cc}\Gamma_{11}&\Gamma_{12}\\
\Gamma_{12}^*&\Gamma_{22}\end{array}\right)\right]
\left(\begin{array}{c}\psi_1\\\psi_2\end{array}\right)~,\label{Eq:7.0}
\Endeq
where $m_{11}$, $m_{22}$, $\Gamma_{11}$, $\Gamma_{22}$ are real and 
$m_{12}$, $\Gamma_{12}$ are complex. The equation was
derived from perturbation theory
in 1930 \cite{1930-WignerWeisskopf} and in a stricter way from 
field theory in 1963 \cite{1963-Sachs}. A comprehensive
treatment of all its properties and symmetries
can be found in \cite{BrancoLavouraSilva}. The phases of
$m_{12}$ and $\Gamma_{12}$ are arbitrary, but their relative phase $\xi$ 
is an observable. CPT invariance requires $m_{11}=m_{22}$ and $\Gamma_{11}=\Gamma_{22}$.
T symmetry requires $\xi=0$ or $\pi$, i.\ e.\ $m_{12}$ and 
$\Gamma_{12}$ can be chosen to be real. CP symmetry requires both CPT and T symmetry;
we write the CP-symmetric equation as
\Begeq
\ri~\frac{\partial}{\partial t }\left(\begin{array}{c}\psi_1\\
\psi_2\end{array}\right)=
\left[\left(\begin{array}{cc}\mu&m_{12}\\m_{12}&\mu\end{array}\right)
-\frac{\ri}{2}\left(\begin{array}{cc}\Gamma&\Gamma_{12}\\
\Gamma_{12}&\Gamma\end{array}\right)\right]
\left(\begin{array}{c}\psi_1\\\psi_2\end{array}\right)~,\label{Eq:7.1}
\Endeq
with four real parameters. 
Its two fundamental solutions are
\bEa
\psi_1+\psi_2 &=& \re^{-(\Gamma+\Gamma_{12})t/2}~\re^{-\ri(\mu+m_{12})t}
~,~~\psi_1-\psi_2 =0~, \nonumber\\
{\rm and}~\psi_1-\psi_2 &=& \re^{-(\Gamma-\Gamma_{12})t/2}~\re^{-\ri(\mu-m_{12})t}
~,~~\psi_1+\psi_2=0~.\label{Eq:7.2}
\eEa
The rate $\Gamma$ is determined by the sum 
of all decays of the $\rMz$. The rate $|\Gamma_{12}|$,
necessarily smaller than $\Gamma$, is determined by all those final
states which are reached from both $\rMz$ and $\rMqz$, examples are 
$\pi^+\pi^-$ and $\pi^0\pi^0$ in the $\rKz\rKqz$ system.\\

\noindent
K and B mesons differ strongly in their ratio
$\Gamma_{12}/\Gamma$. For the Kaons, the sum of $\pi^+\pi^-$ and 
$\pi^0\pi^0$ dominates all decays; $|\Gamma_{12}|=\Gamma$ in good
approximation. We choose the unobservable sign of $\Gamma_{12}$ to be
negative,
\Begeq
\Gamma(\rK)+\Gamma_{12}(\rK)\approx 0~,~~\Gamma(\rK)-\Gamma_{12}(\rK)
\approx 2 \Gamma(\rK)~,
\Endeq
and $\psi_1-\psi_2$ in Eq.~\ref{Eq:7.2} describes the $\rKz_\rS$,
$\psi_1+\psi_2$ the $\rKz_\rL$. In the B-meson system, $\Gamma$ is 
dominated by those final states which are
reached from either only $\rBz$ or only $\rBqz$. In contrast to
$|\Gamma_{12}(\rK)/\Gamma(\rK)|\approx 1$, we have
$|\Gamma_{12}(\rB)/\Gamma(\rB)|\ll 1$ and for the demonstation we
can approximate $\Gamma_{12}(\rB)=0$. For the D-meson pair we have
$|\Gamma_{12}(\rD)/\Gamma(\rD)|\ll 1$ as well, the distinction is
the smallness of $|m_{12}(\rD)|$. 
The ratio $|m_{12}(\rD)/\Gamma(\rD)|$ is of the order $10^{-2}$, whereas 
$|m_{12}(\rB)/\Gamma(\rB)|=o(1)$.\\
 
\noindent
Setting $m_{12}=-k$, $\mu+m_{12}=m$ and using the above approximatons for
$\Gamma_{12}$, we obtain for the K mesons
\bEa
\left(\begin{array}{c}\ddot\Re_1\\ \ddot\Re_2\end{array}\right)
=-\left(\begin{array}{cc}m^2+2mk+2k^2+\Gamma^2/2&-2mk-2k^2-\Gamma^2/2\\ 
-2mk-2k^2-\Gamma^2/2&m^2+2mk+2k^2+\Gamma^2/2\end{array}\right)
\left(\begin{array}{c}\Re_1\\ \Re_2\end{array}\right)\nonumber\\
-\left(\begin{array}{cc}\Gamma&-\Gamma\\-\Gamma &\Gamma\end{array}\right)
\left(\begin{array}{c}\dot\Re_1\\ \dot\Re_2\end{array}\right)~,
\label{Eq:7.1R}\eEa
and for the B and D mesons
\bEa
\left(\begin{array}{c}\ddot\Re_1\\ \ddot\Re_2\end{array}\right)
=-\left(\begin{array}{cc}m^2+\Gamma^2/4+2mk+2k^2&-2mk-2k^2\\ 
-2mk-2k^2&m^2+\Gamma^2/4+2mk+2k^2\end{array}\right)
\left(\begin{array}{c}\Re_1\\ \Re_2\end{array}\right)
\nonumber\\
-\left(\begin{array}{cc}\Gamma&0\\0&\Gamma\end{array}\right)
\left(\begin{array}{c}\dot\Re_1\\ \dot\Re_2\end{array}\right)~.
\eEa
We now continue with the equations for the corresponding motions
of the pendula.\\[3mm]
 
\noindent
6. Pendulum Motions demonstrating the three Transitions $\rMz
\leftrightarrow\rMqz$\\

\begin{figure}[h]
  \centering
  \begin{minipage}[ ]{7cm}
  \hfill\includegraphics[width=6cm]{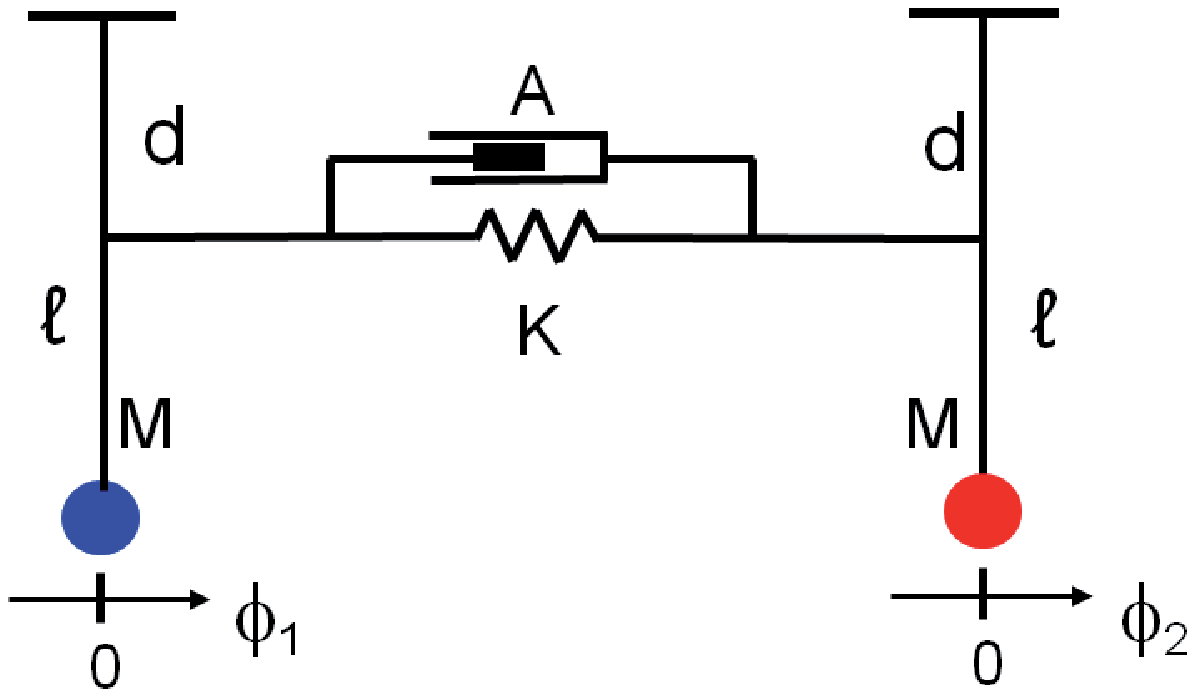}\hfill
  \caption[ ]{``$\rKz\rKqz$ pendula" with damping in the coupling.
  $A$ is the proportionality constant between $F_{12}$ and 
  $d\cdot (\dot\phi_2-\dot\phi_1)$,
  the other  notations are the same as in Fig.~\ref{Fig:6.1}.
  \label{Fig:8.1}}\end{minipage}\hfill
  \begin{minipage}[ ]{7.5cm}
  \includegraphics[width=7.2cm]{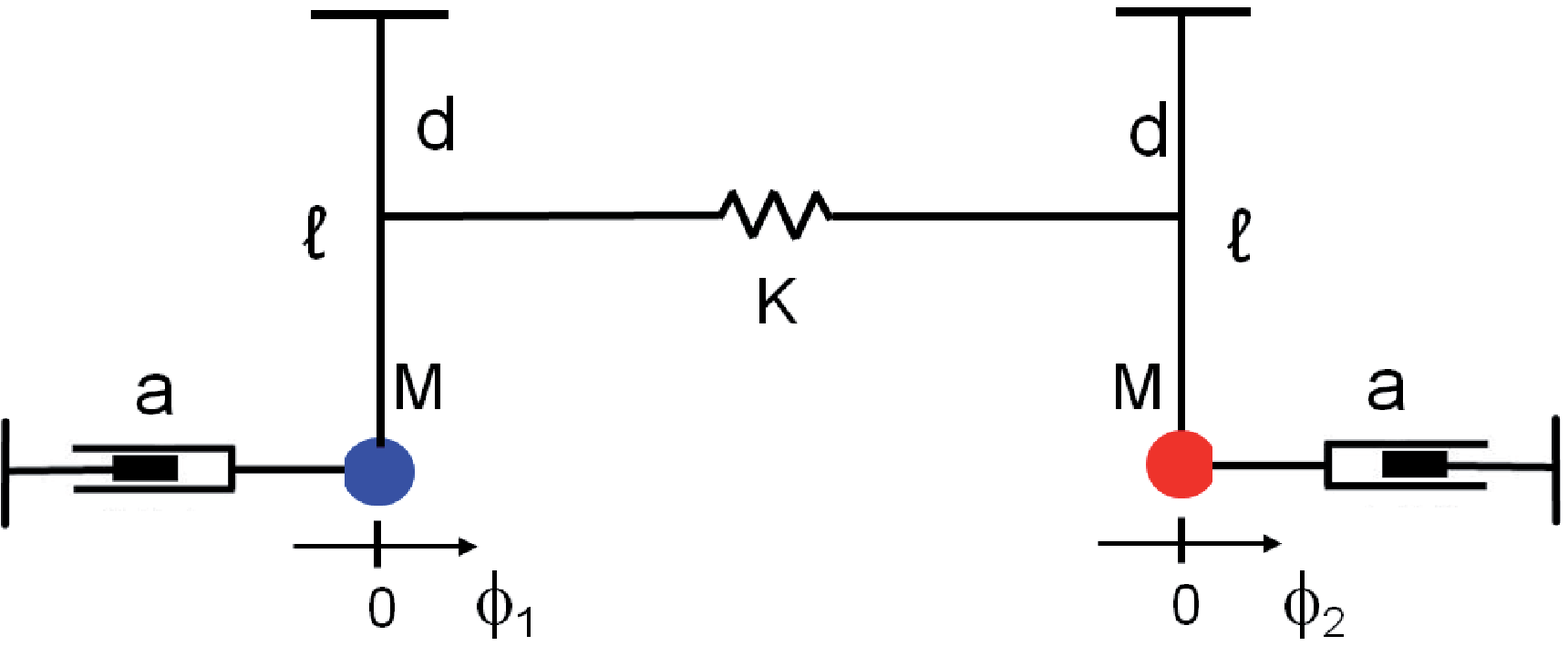}
  \caption[ ]{``$\rBz\rBqz$ and $\rDz\rDqz$ pendula" with separate 
  dampings for each pendulum. The constant $a$ is the proportionality 
  between damping force and $\ell\cdot \phi_\ri$. 
  \label{Fig:9.1}}\end{minipage}
\end{figure}

\noindent
The velocity-proportional dampings applied to the pendula
are shown in Figs.~\ref{Fig:8.1} and \ref{Fig:9.1}. For the
K system, the coupling force is
\Begeq
F_{12}=-F_{21}=-Kd(\phi_1 - \phi_2)-Ad(\dot\phi_1 -\dot \phi_2)
\Endeq
and the equation of motion becomes
\bEa
M\ell^2\left(\begin{array}{c}\ddot\phi_1\\ \ddot\phi_2\end{array}\right)
=-\left(\begin{array}{cc}Mg\ell+Kd^2&-Kd^2\\ 
-Kd^2&Mg\ell+Kd^2\end{array}\right)
\left(\begin{array}{c}\phi_1\\ \phi_2\end{array}\right)\nonumber\\
-\left(\begin{array}{cc}Ad^2&-Ad^2\\ -Ad^2&Ad^2\end{array}\right)
\left(\begin{array}{c}\dot\phi_1\\ \dot\phi_2\end{array}\right)~.
\label{Eq:8.2}
\eEa
With $\Gamma=Ad^2/(M\ell^2)$ this is in agreement with
Eq.~\ref{Eq:7.1R}.
Video07 \cite{links} shows the undamped parallel fundamental 
solution of 
Eq.~\ref{Eq:8.2} representing the state $\rKz_\rL$, 
Video08 \cite{links} the damped antiparallel motion representing 
the $\rKz_\rS$. In our setup,
the damping time $1/2\Gamma$ is 7 s. If we start the
movement with the state $\rKqz$, see Video09 \cite{links},
its $\rKz_\rS$ component is damped away quickly; at large times we observe
only its undamped component $\rKz_\rL$. A completely symmetric movement
is seen in Video10 \cite{links} when we start with the 
right-side pendulum $\rKz$.\\

\noindent
For demonstrating $\rBz\rBqz$ transitions, we change the setup according
to Fig.~\ref{Fig:9.1} by removing the damping in the coupling and 
using the two eddy-current brakes of equal strength.
The damping forces are now $-a\ell\phi_\ri$ (\ri=1,2), 
the damping torques $-a\ell^2\phi_\ri$, and the equation of motion is
\Begeq
M\ell^2\left(\begin{array}{c}\ddot\phi_1\\ \ddot\phi_2\end{array}\right)
=-\left(\begin{array}{cc}Mg\ell+Kd^2&-Kd^2\\ 
-Kd^2&Mg\ell+Kd^2\end{array}\right)
\left(\begin{array}{c}\phi_1\\ \phi_2\end{array}\right)
-\left(\begin{array}{cc}a\ell^2&0\\ 0&a\ell^2\end{array}\right)
\left(\begin{array}{c}\dot\phi_1\\ \dot\phi_2\end{array}\right)\ .
\label{Eq:9.2}
\Endeq
\vspace{2mm}

\noindent
Video11 \cite{links} and Video12 \cite{links} show the 
fundamental modes; both 
have the same lifetime, equal to that of the blue and the red mode.
In Video13 \cite{links} and Video14 \cite{links}, 
where we start the motion with the 
blue and the red pendulum respectively, we show the time evolution of the 
$\rBz\rBqz$ transitions.
Coupling and dampings are chosen in such a way that the observed 
transition strength $\chi(\rB)=0.19$ is approximated.\\
 
\noindent
For demonstrating $\rDz\rDqz$ transitions, we change ony the coupling
weight and keep the same individual dampings for the two pendula.
The lifetimes $1/\Gamma(\rD)$ and $1/\Gamma(\rB)$ are of the same order,
but the coupling $|m_{12}(\rD)|$ is two orders of magnitude smaller than
$|m_{12}(\rB)|$. Video15
\cite{links} shows this case with the red pendulum starting. Its motion
is nearly damped away before there is a visible transition of motion 
energy from the red to the blue side.\\[3mm]

\noindent
7. CP Violation in $\rKz\rKqz$ Transitions\\

\noindent
As shown in Eq.~\ref{Eq:7.2}, CP symmetry in $\rKz\rKqz$ transitions leads
to two fundamental modes with the properties $\psi_1-\psi_2=0$ and
$\psi_1+\psi_2=0$, respectively. If the symmetry is broken this is no
longer the case, and the mode $\rKz_\rL$ which is reached from $\rKz$ or
$\rKqz$ for times $t\gg 1/\Gamma_\rS$ contains different fractions of
$\rKz$ and $\rKqz$,
\Begeq
\rKz_\rL=p~\rKz+q~\rKqz~{\rm with}~|p/q|\ne 1~.
\label{Eq:11.3}
\Endeq
The experimental result is $|p/q| = 1.00332 \pm 0.00003$ \cite{2010-PDG}.\\
 
\begin{figure}[h]\begin{center}
\includegraphics[width=5cm]{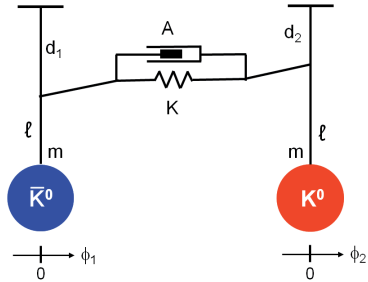}\end{center}
\caption[ ]{Asymmetric lever arms for the coupling.\label{Fig:12.1}}
\end{figure}

\noindent
For a simulation with the pendula, we use the setup in Fig.~\ref{Fig:12.1}.
The coupling forces
have to obey $F_{21}=-F_{12}$, but the torques can be different if
we choose two different lever arms $d_1>d_2$.
Using $d_1 = 20.5$ cm, $d_2 = 15.5$ cm, $\ell\phi_1 (0) =$
10 cm, and $\ell\phi_2 (0) =0$ we show in Video16 
\cite{links} that,
after damping the $\rKz_\rS$ mode, the amplitude of the red $\rKz$ in the
surviving $\rKz_\rL$ is larger than that of the blue $\rKqz$. The former is
5~cm, the latter 4~cm. Video17 \cite{links} shows the same larger
amplitude of the $\rKz$ if we start the motion with the other-side
pendulum. These two motions in Video16 and Video17 represent the
observed CP violation in the $\rKz_\rL$ state.

\noindent
The equation of motion for this setup is
\bEa
M\ell^2\left(\begin{array}{c}\ddot\phi_1\\ \ddot\phi_2\end{array}\right)
=-\left(\begin{array}{cc}Mg\ell+Kd_1^2&-Kd_1 d_2\\ 
-Kd_1 d_2&Mg\ell+Kd_2^2\end{array}\right)
\left(\begin{array}{c}\phi_1\\ \phi_2\end{array}\right)\nonumber\\
-\left(\begin{array}{cc}Ad_1^2&-Ad_1 d_2\\ -Ad_1 d_2&Ad_2^2\end{array}
\right)\left(\begin{array}{c}\dot\phi_1\\ \dot\phi_2\end{array}\right)~.
\label{Eq:12.4}
\eEa
The symmetries of the two matrices (equal off-diagonal and different
diagonal elements) are different from those originating from a CPT-symmetric 
Schr\"o\-din\-ger equation, as shown in Appendix 2.
$\rKz\rKqz$ transitions, however, break both CP and T symmetry
and no violation of CPT has been found
\cite{1970-Schubert,1998-CPLEAR,2006-FidecaroGerber}. 
In fact, no friction-damped pendulum pair with oscillations in the same 
space dimension
can have an equation of motion originating from a CPT-symmetric and 
T-violating Schr\"odinger equation
\cite{1988-Winstein,2001-KosteleckyRoberts}.
To our judgement, this should not hinder us to include the present chapter
into the article because one of the main features of CP violation in
$\rKz\rKqz$ transitions is clearly demonstrated.\\

\noindent 
Examples for coupled
and damped oscillations with a CPT-symmetric and T-violating equation
of motion are given in the literature; a Foucault pendulum with
damping in one direction \cite{1996-Rosner,2001-RosnerSlezak}, 
a ball rolling in a bowl on a turntable \cite{2001-KosteleckyRoberts},
and an electrical setup in which the 
non-reciprocity
in the coupling between two resonant circuits is achieved by a gyrator
\cite{2001-KosteleckyRoberts,2011-Caruso}.
To our knowledge, none of the proposed demonstration setups has been built so far.  
\\[3mm]

\noindent
Summary\\

\noindent
Starting from two-state Schr\"odinger equations for pairs of 
coupled mesons, we derive second-order differential
equations for the real parts of their solutions and map them 
onto equations of motion for a macro-mechanical system. 
This is a pair of coupled identical pendula with a possible damping 
in the coupling and individual dampings for each of the pendula.
By adjusting the damping and coupling parameters, three different
particle-antiparticle transitions and the effect of CP violation can 
be visualized. The paper is accompanied by a sequence of video clips 
\cite{links} that
demonstrate the performance of the pendula. The case of 
CPT symmetry along with CP and T violation cannot be reproduced with the setup 
described. Detailed construction drawings are attached \cite{links} so that the 
interested reader can build his/her own pendulum pair in order to demonstrate
the correspondence between quantum mechanics and macroscopic physics.\\[3mm]

\noindent
Appendix 1. Why does Re($\Psi$) carry the same information as $\Psi$?\\

\noindent
The most general solution of the linear two-component 
Schr\"odinger equation in Eq.~\ref{Eq:7.0} is
\Begeq
\left(\begin{array}{c}\psi_1\\ \psi_2\end{array}\right) = 
a\cdot \left(\begin{array}{c}p\\ q\end{array}\right)
\cdot\re^{-\ri m_St}~\re^{-\Gamma_St/2}+b\cdot \re^{\ri\phi}\cdot 
\left(\begin{array}{c}r\cdot\re^{\ri\delta}\\ s\end{array}\right)
\cdot\re^{-\ri m_Lt}~\re^{-\Gamma_Lt/2}~.\label{Eq:A2.1}                        
\Endeq
With $p^2+q^2=1$ and $r^2+s^2=1$ it contains 10 real parameters, 
7 given by the equation
[$m=(m_S+m_L)/2,~\Delta m=m_S-m_L,~\Gamma=(\Gamma_S+\Gamma_L)/2,~
\Delta\Gamma=\Gamma_S-\Gamma_L,~p/q,~r/s,~\delta$] 
and 3 by the initial conditions
$[a,~b,~\phi]$. Without CPT symmetry, $r/s \ne -p/q$ and the phase
$\delta$ is observable. For the following we also need the
weak-interaction conditions $\Gamma\ll m$ and
$|\Delta m|\ll m$. 
The real part of $\psi_1$ is
\Begeq
\Re_1=a\cdot p\cdot\cos(mt+\Delta m\cdot t/2)~\re^{-\Gamma_St/2}+b\cdot r 
\cdot\cos(mt-\Delta m\cdot t/2 +\phi+\delta)~\re^{-\Gamma_Lt/2}~.
\Endeq
With the weak-interaction conditions we obtain
\Begeq
\langle \cos^2m_St\rangle=\langle \cos^2m_Lt\rangle = \langle \cos^2mt\rangle
=\langle \sin^2mt\rangle =1/2~,~~\langle \cos mt\cdot\sin mt\rangle=0~,
\Endeq
\Begeq
2\langle \Re_1^2\rangle=a^2p^2\cdot\re^{-\Gamma_St}+b^2r^2\cdot\re^{-\Gamma_Lt}+
2~abpr\cdot\re^{-\Gamma t}\cos(\Delta m\cdot t+\phi+\delta)~.
\Endeq
The calculation of $|\psi_1|^2$ gives exactly the same result. The
conclusion $2\langle \Re_2^2\rangle=|\psi_2|^2$ is found by just
replacing $p\to q$, $r\to s$, and $\phi+\delta\to\phi$.\\

\noindent
The equivalence of $\psi$ and Re($\psi$) is far from being general. A
simple counter-example is the harmonic oscillator with strong damping.
For $\psi=\re^{-\ri mt}\cdot\re^{-\Gamma t/2}$ with $\Gamma/m=o(1)$ we cannot find
averaging intervals with 
$\rm\langle Re^2(\psi)\rangle =\langle Im^2(\psi)\rangle$.\\[3mm]

\noindent
Appendix 2. CPT Symmetry with CP and T Violation\\

\noindent
The Schr\"odinger equation for CPT-conserving but CP- and T-violating
$\rMz\rMqz$ transitions is
\Begeq 
\ri\frac{\partial}{\partial t }\left(\begin{array}{c}\psi_1\\
\psi_2\end{array}\right)=
\left[\left(\begin{array}{cc}\mu&m_{12}\\m_{12}^*&\mu\end{array}\right)
-\frac{\ri}{2}\left(\begin{array}{cc}\Gamma&\Gamma_{12}\\
\Gamma_{12}^*&\Gamma\end{array}\right)\right]
\left(\begin{array}{c}\psi_1\\\psi_2\end{array}\right)~;\label{Eq:A.1} 
\Endeq
with five observable real parameters, $\mu$, $\Gamma$, $|m_{12}|$,
${\rm Re}~(\Gamma_{12}/m_{12})$, and  ${\rm Im}~(\Gamma_{12}/m_{12})$.
Its two fundamental solutions are
\Begeq
\left(\begin{array}{c}\psi_1\\ \psi_2\end{array}\right)_S = 
\left(\begin{array}{c}p\\ q\end{array}\right)
\cdot\re^{-\ri m_S t}~\re^{-\Gamma_St/2}~,~~
\left(\begin{array}{c}\psi_1\\ \psi_2\end{array}\right)_L = 
\left(\begin{array}{c}p\\ -q\end{array}\right)
\cdot\re^{-\ri m_Lt}~\re^{-\Gamma_Lt/2}~.\label{Eq:A.2}                        
\Endeq
The values of the five observables $m_S$, $\Gamma_S$, $m_L$, $\Gamma_L$,
and $|p/q|$ follow unambiguously from the five parameters of the
Schr\"odinger equation \cite{BrancoLavouraSilva} . The values of $p$ and
$q$ can be chosen to be real, leading to
\Begeq
\left(\begin{array}{c}\Re_1\\ \Re_2\end{array}\right)_S = 
\left(\begin{array}{c}p\\ q\end{array}\right)
\cdot\re^{-\Gamma_St/2}~\cos(m_St)~,~~
\left(\begin{array}{c}\Re_1\\ \Re_2\end{array}\right)_L = 
\left(\begin{array}{c}p\\ -q\end{array}\right)
\cdot\re^{-\Gamma_Lt/2}~\cos(m_Lt)~.                        
\Endeq
What is the differential equation for the most general superposition,
\Begeq
\Re_1~|\rMz\rangle+\Re_2~|\rMqz\rangle=
\left(\frac{\Re_1}{2p}+\frac{\Re_2}{2q}\right)~|\rMz_S\rangle~+~
\left(\frac{\Re_1}{2p}-\frac{\Re_2}{2q}\right)~|\rMz_L\rangle~?                        
\Endeq
The $\rMz_S$ and $\rMz_L$ components obey the differential equations
\bEa
\left(\frac{\ddot\Re_1}{2p}+\frac{\ddot\Re_2}{2q}\right)~=
-(m_S^2+\Gamma_S^2/4)~\left(\frac{\Re_1}{2p}+\frac{\Re_2}{2q}\right)
-\Gamma_S~\left(\frac{\dot\Re_1}{2p}+\frac{\dot\Re_2}{2q}\right)~,
\nonumber\\
\left(\frac{\ddot\Re_1}{2p}-\frac{\ddot\Re_2}{2q}\right)~=
-(m_L^2+\Gamma_L^2/4)~\left(\frac{\Re_1}{2p}-\frac{\Re_2}{2q}\right)
-\Gamma_L~\left(\frac{\dot\Re_1}{2p}-\frac{\dot\Re_2}{2q}\right)~,
\label{Eq:I.3}
\eEa
resulting in
\bEa
\left(\begin{array}{c}\ddot\Re_1\\ \ddot\Re_2\end{array}\right)=-\left(
\begin{array}{cc}\frac{m_S^2+m_L^2}{2}+\frac{\Gamma_S^2+\Gamma_L^2}{8}&
\frac{p}{q}\cdot\left(\frac{m_S^2-m_L^2}{2}+\frac{\Gamma_S^2-\Gamma_L^2}{8}
\right)\\\frac{q}{p}\cdot
\left(\frac{m_S^2-m_L^2}{2}+\frac{\Gamma_S^2-\Gamma_L^2}{8}\right)&
\frac{m_S^2+m_L^2}{2}+\frac{\Gamma_S^2+\Gamma_L^2}{8}\end{array}\right)
\left(\begin{array}{c}\Re_1\\ \Re_2\end{array}\right) \nonumber \\
-\left(\begin{array}{cc}\frac{\Gamma_S+\Gamma_L}{2}&
\frac{p}{q}\cdot\frac{\Gamma_S-\Gamma_L}{2}\\
\frac{q}{p}\cdot\frac{\Gamma_S-\Gamma_L}{2}&\frac{\Gamma_S+\Gamma_L}{2}
\end{array}\right)\left(\begin{array}{c}\dot\Re_1\\ \dot\Re_2
\end{array}\right)\ .\label{Eq:I.4}
\eEa
With CP and T violation, i.\ e.\
$p/q\ne 1$, the differential equation for the real parts exhibits the
same symmetry properties as the Schr\"odinger equation; the diagonal matrix
elements are equal, the off-diagonal elements 12 are different.\\[3mm]

\noindent
Acknowledgments\\

\noindent
We are grateful to Rolf Musch and Helmut Maier in the mechanical workshop of 
the former Institut f\"ur Hochenergiephysik at Heidelberg for the 
construction of our pendulum pair, and to Christian Herdt for the
construction drawing. We also thank Michael Kobel, Karlheinz Meier,
and Otto Nachtmann
for support and fruitful discussions, Hans-Georg Siebig, Harald
Schubert, and Daniel Stegen for their help 
in producing the video clips, and Ernest Henley, Andrzej Buras, Svjetlana Fajfer,
and Barbara Schrempp  for their
encouragements to publish this article.

\end{document}